\documentclass[useAMS,usenatbib]{mn2e}
\usepackage{times}
\usepackage{graphicx}
\usepackage{subfigure}
\usepackage{psfrag}
\usepackage{amsmath}
\usepackage{amsfonts}
\usepackage{amssymb} 
\usepackage{flafter} 

\def\reff@jnl#1{{\rm#1\/}}
\def\aj{\reff@jnl{AJ}}                 
\def\araa{\reff@jnl{ARA\&A}}           
\def\apj{\reff@jnl{ApJ}}               
\def\apjl{\reff@jnl{ApJ}}              
\def\apjs{\reff@jnl{ApJS}}             
\def\ao{\reff@jnl{Appl.Optics}}        
\def\apss{\reff@jnl{Ap\&SS}}           
\def\aap{\reff@jnl{A\&A}}              
\def\aapr{\reff@jnl{A\&A~Rev.}}        
\def\aaps{\reff@jnl{A\&AS}}            
\def\azh{\reff@jnl{AZh}}               
\def\baas{\reff@jnl{BAAS}}             
\def\jrasc{\reff@jnl{JRASC}}           
\def\memras{\reff@jnl{MmRAS}}          
\def\mnras{\reff@jnl{MNRAS}}           
\def\pra{\reff@jnl{Phys.Rev.A}}        
\def\prb{\reff@jnl{Phys.Rev.B}}        
\def\prc{\reff@jnl{Phys.Rev.C}}        
\def\prd{\reff@jnl{Phys.Rev.D}}        
\def\prl{\reff@jnl{Phys.Rev.Lett}}     
\def\pasp{\reff@jnl{PASP}}             
\def\pasj{\reff@jnl{PASJ}}             
\def\qjras{\reff@jnl{QJRAS}}           
\def\skytel{\reff@jnl{S\&T}}           
\def\solphys{\reff@jnl{Solar~Phys.}}   
\def\sovast{\reff@jnl{Soviet~Ast.}}    
\def\ssr{\reff@jnl{Space~Sci.Rev.}}    
\def\zap{\reff@jnl{ZAp}}               
\def\nat{\reff@jnl{Nature}}            

\newcommand{\grad}{\nabla}
\newcommand{\half}{{\textstyle \frac{1}{2}}}
\newcommand{\bgrad}{\mbox{\boldmath $\grad$}}
\newcommand{\br}{\mbox{\boldmath $r$}}

\newcommand{\be}{\begin{equation}}
\newcommand{\ee}{\end{equation}}
\newcommand{\barr}{\begin{array}}
\newcommand{\earr}{\end{array}}
\newcommand{\bea}{\begin{eqnarray}}
\newcommand{\eea}{\end{eqnarray}}
\newcommand{\beqa}{\be \begin{array}{rcl}}
\newcommand{\eeqa}{\end{array} \ee}

\newcommand{\dt}{{\cdot}}

\newcommand{\bv}{\mbox{\boldmath $v$}}

\newcommand{\schro}{Schr\"odinger\;}

\title[\schro fluid dynamics]
{Cosmological fluid dynamics in the \schro formalism}
\author[Rebecca Johnston, A.N.~Lasenby and M.P.~Hobson]
{Rebecca Johnston\thanks{e-mail: rrj20@mrao.cam.ac.uk},
A.N.~Lasenby and M.P.~Hobson\\
Astrophysics Group, Cavendish Laboratory, J.J.~Thomson Avenue,
Cambridge CB3 0HE, UK}

\begin{document}

\date{Accepted ---. Received ---; in original form \today}
\pagerange{\pageref{firstpage}--\pageref{lastpage}}
\pubyear{2009}

\voffset=-0.8in

\label{firstpage}
\maketitle

\begin{abstract}
\noindent 
We investigate the dynamics of a cosmological dark matter fluid in the
Schr\"odinger formulation, seeking to evaluate the approach as a
potential tool for theorists. We find simple wave-mechanical solutions
of the equations for the cosmological homogeneous background evolution
of the dark matter field, and use them to obtain a piecewise analytic
solution for the evolution of a compensated spherical overdensity.  We
analyse this solution from a `quantum mechanical' viewpoint, and
establish the correct boundary conditions satisfied by the
wavefunction. Using techniques from multi-particle quantum mechanics,
we establish the equations governing the evolution of multiple
fluids and then solve them numerically in such a system.  Our results
establish the viability of the \schro formulation as a genuine
alternative to standard methods in certain contexts, and a novel way
to model multiple fluids.
\end{abstract}

\begin{keywords}
cosmology:theory -- galaxies: clusters: general
\end{keywords}

\section{Introduction}\label{sec:intro}

On scales smaller than the Hubble distance, the evolution of structure
in the universe can be analysed using Newtonian equations for
gravitational instability (Jeans 1928). In particular, the phase space
evolution of the dominant cold dark matter (CDM) component is governed
by a collisionless Boltzmann, or Vlasov, equation, coupled to a
Poisson equation for gravity. When velocity dispersion is negligible,
a fluid approximation holds, which leads to a much simpler
representation, but one that fails to accommodate multiple streams in
the fluid, most notably at the onset of so-called `shell-crossing' in
the non-linear evolution of density perturbations.

In the linear regime, one can (by definition) describe structure
growth accurately using first-order perturbation theory (Peebles 1980)
applied to the standard Newtonian equations for a self-gravitating
fluid. In the strongly nonlinear regime, only numerical simulations
are capable of modelling general fluid evolution and reproducing the
structures observed in galaxy surveys; for a recent review, see
Springel et al. 2006. However, the evolution of density fluctuations
in the so-called `weakly nonlinear regime', those characteristically
represented on the largest scales in galaxy clustering surveys, have
been well described by a variety of semi-analytic methods. These
include nonlinear perturbation theory, and related schemes such as the
Zeldovich and adhesion approximations (see Bernardeau et al. 2002 for
a review). These methods typically extrapolate results from
the linear regime into the quasi-linear regime.

In a separate line of research, the correspondence between the
equations describing a collisionless fluid and the \schro equation was
first identified by Madelung in 1926. Outside cosmology, this
correspondence is most commonly used in drawing a hydrodynamic analogy
for quantum mechanics (Spiegel 1980; Skodje et al. 1989). Widrow \&
Kaiser (1993) were the first to apply the ideas to the problem of
cosmological structure formation, using a \schro field representation
for collisionless matter, but omitting the `quantum pressure' term to
obtain a linear wave-equation. They developed a numerical technique to
follow the nonlinear evolution of the field, offering an
alternative to N-body particle mesh codes. A fuller explanation and
exploration of their method is given by Davies \& Widrow (1997).

Coles \& Spencer (2002) take a different approach, using the
so-called Madelung transformation to establish a simpler
correspondence between the self-gravitating fluid equations and a
Schr\"odinger--Poisson system. This transformation means that their
representation of the wavefunction can only be used for single
streaming fluids, with their aim to develop a semi-analytic technique
for tracking the development of structure into the quasilinear
regime. Short \& Coles (2006a,b) develop this foundational work into
what they term the free-particle approximation for precisely this
purpose, in which the gravitational potential and the quantum pressure
are neglected.  They show that this approximation is useful into the
mildly nonlinear regime and the resulting method is essentially
equivalent to the well-known adhesion approximation. Szapudi \&
Kaiser (2003) develop a \schro perturbation theory very similar in
nature to the well-known perturbation theory of the standard Eulerian
fluid equations.

With the exception of Szapudi \& Kaiser, all previous work in this
area has been aimed chiefly at developing better computational
techniques for tracking the evolution of structure, based almost
incidentally on a wave mechanical formulation, which may confer
certain advantages over rival methods; the full wave-mechanical system
of equations that govern structure growth has not been a feature of
interest. In this paper, our aim is to explore the \schro 
formulation of the evolution of structure from a fully self-consistent
point of view, without recourse to approximations.

The plan of the paper is as follows. In Section~\ref{sec:ourapproach},
we outline the \schro formalism for fluid dynamics, summarizing the
previous work in the field and outlining our own direct approach.  In
Section~\ref{sec:cosmosolns}, we obtain solutions to the
Schr\"odinger--Poisson system for spatially-flat and closed
homogeneous cosmological models. These solutions are then used in
Section~\ref{sec:sphericalmodel} to derive a wave-mechanical solution
for a compensated spherical overdensity model. In
Section~\ref{sec:shellcrossing}, we describe an extension to the
\schro formalism, based on techniques from multi-particle quantum
mechanics, that allows one to accommodate multiple fluids. We
apply this technique to a modified spherical overdensity model that
exhibits multi-streaming by integrating the evolution equations
numerically. Finally, our conclusions are presented in
Section~\ref{sec:conc}

\section{Schr\"odinger formalism for fluids}\label{sec:ourapproach}

We take as our starting point the usual Newtonian equations for
a self-gravitating perfect fluid.
The continuity, Euler and Poisson equations comprise the first
and second momentum moments of the full Vlasov--Poisson system, and
are exact provided the velocity dispersion is zero, i.e. for single
streaming fluids. These equations are:
\begin{eqnarray}
\label{Continuity}
   \frac{\partial \rho}{\partial t} + \nabla\cdot(\rho \bmath{v}) & = & 0, \\
\label{Euler}
    \frac{\partial \bmath{v}}{\partial t} + (\bmath{v}\cdot\nabla)\bmath{v}
    +\frac{1}{\rho}\nabla p + \nabla V & = & 0, \\
\label{PoissonFluid}
  \nabla^{2} V  & = & 4 \pi G \rho - \Lambda c^2.
\end{eqnarray}
Here $\bmath{v}$, $\rho$ and $p$ are the fluid velocity, density and
pressure respectively, and $V$ is the gravitational potential.  In
(\ref{PoissonFluid}), we have modified the Poisson equation to include
a term representing the repulsive force provided by a non-zero
cosmological constant $\Lambda$.

Adopting the usual description of cold dark matter as a pressureless
fluid, and restricting our analysis to irrotational flow, which is
normally considered for cosmological evolution in homogeneous and
isotropic universe models, one can replace the Euler equation
(\ref{Euler}) with the Bernoulli equation
\begin{equation}
\frac{\partial\phi}{\partial
  t}+\tfrac{1}{2}\left(\nabla\phi\right)^2=-V,
\label{bernoulli}
\end{equation}
with $\bmath{v} = \nabla \phi$ where $\phi$ is the velocity potential.

Since our equations describe a single streaming fluid, one may
introduce the Madelung transformation
\begin{equation}
\label{wavefunctiondefinition}
\psi = \alpha e^{i \phi / \nu}.
\end{equation}
In this transformation, $\rho = |\psi|^2 = \alpha^2$, so the fluid
density is always nonnegative, and $\nu$ is an adjustable parameter
which has the same dimension as $\phi$, namely $[L^2 T^{-1}]$.  It is
then easy to show that the equations governing the evolution of $\psi$
and $V$ are
\begin{eqnarray}
\label{spsystem1}
  i\nu \frac{\partial\psi}{\partial t} &=& \frac{\nu^{2}}{2}\nabla^{2}\psi + V\psi +\frac{\nu^{2}}{2}\frac{\nabla^{2}| \psi |}{|\psi |} \psi, \\
\label{spsystem2}
  \nabla^{2}V &=& 4\pi G |\psi | ^{2} - \Lambda c^2,
\end{eqnarray}

Thus, we see that the dynamics of a dark matter fluid may be described
by the (modified) Poisson equation coupled to a nonlinear wave
equation.  This permits a direct formal
correspondence to be made between the concepts of fluid dynamics and
those of quantum mechanics, leading to potential conceptual advantages
such as fluid quantities being associated with Hermitian
operators. There are also some possible computational
simplifications. In particular, the WKB limit for the wave-equation
corresponds to the asymptotic study of the limit $\nu \to 0$ (Spiegel
1980).

Nonetheless, the nonlinear nature of the wave-equation
(\ref{spsystem1}) does present some conceptual difficulties.
Nonlinear Schrodinger equations and their applications in
(non-cosmological) physics are treated thoroughly in Sulem \& Sulem
(1999) and Pang \& Feng (2005). The theory of nonlinear quantum
mechanics is used to describe macroscopic quantum effects and
quasi-particles such as solitons, which have a Hamiltonian which
depends nonlinearly on their wavefunction. Typically used in the
modelling of quantum semiconductors and superfluids, the theory
differs from linear quantum mechanics in many important ways. For
example, the square of the wavefunction is no longer the probability
of finding the macroscopic particle at a given point in spacetime, but
instead gives the mass density of its constituent microscopic
particles at that point. Operators are no longer linear, and hence
wavefunctions no longer obey the principle of superposition; the
unitary structure of linear quantum mechanics does not apply.

In addition to these conceptual differences with standard quantum
mechanics, the nonlinear nature of the coupled Schr\"odinger--Poisson
equations (\ref{spsystem1}) and (\ref{spsystem2}) make them difficult
to solve. The wave-equation has two sources of nonlinearity: the first
from the coupling to the Poisson equation via the gravitational
potential, and the second from the nonlinear `quantum pressure' term
(this term resembles a pressure gradient when interpreting quantum
phenomena in terms of classical fluid behavior)
\begin{equation}
P = \frac{\nu^2}{2}\frac{\nabla^{2}| \psi |}{| \psi |} \psi,
\label{qpdef}
\end{equation}
which acts like an effective potential. It is useful to insert a
multiplicative parameter $\beta$ in the nonlinear quantum pressure
term in the wave-equation (\ref{spsystem1}), such that $\beta = 0$
corresponds to leaving the term out, and $\beta=1$ to retaining
it. One then finds that the corresponding fluid equations remain
unaltered, except for the Bernoulli equation (\ref{bernoulli}), which
now takes the form
\begin{equation}
\frac{\partial\phi}{\partial t}
+\tfrac{1}{2}\left(\nabla\phi\right)^2=-V+\left(1-\beta\right)P.
\label{bernoulli2}
\end{equation}

The previous studies mentioned thus far, which apply the \schro
formalism to the dynamics of a cosmological dark matter fluid, 
typically discard the troublesome nonlinear quantum pressure term
$P$. From (\ref{qpdef}), we see that omitting $P$ in the
modelling of a dark matter fluid can be justified only if
$|\psi|=\sqrt{\rho}$ varies very slowly on the scales of
interest. Alternatively, for fluids with pressure (such as a baryon
component), one could adjust (or omit) the $P$ term in the wave
equation to model genuine effects of pressure (Coles \& Spencer 2002),
but this is clearly not relevant to modelling a dark matter fluid.

We note from (\ref{qpdef}), however, that in the limit $\nu \to 0$,
equation (\ref{bernoulli2}) reduces to the correct Bernoulli equation
(\ref{bernoulli}). This has led several authors to discard the quantum
pressure term $P$, thereby making the resulting wave-equation linear,
and considering the resulting theory only in the `correspondence limit'
$\nu \to 0$. Although, in this limit, the fluid equations to which
this modified Schr\"odinger--Poisson system is equivalent are those
that hold physically, the limit $\nu \to 0$ is nonetheless meaningless
in the context of the Schr\"odinger formulation itself; moreover, this
assumption constrains these previous authors to considering the study
of phenomena in a certain limit of a parameter in their theory. 
Short \& Coles, for example, tackle this difficulty by performing
simulations for the lowest value of $\nu$ possible numerically and,
unsurprisingly, find their simulations agree best with $N$-body methods
for smallest possible $\nu$. Szapudi \& Kaiser's perturbation theory
takes a more rigorously analytic approach, but it is nonetheless based
on a linearised \schro equation, and thus is only valid in the
correspondence limit. Furthermore, in developing their free particle
approximation, Short \& Coles rely heavily on results from linear
perturbation theory, which is an entirely different mathematical
representation of the physical process of interest. In fact, no study
yet exists of the mathematical formulation of structure formation in
the Schr\"odinger form which is not approximative in some way.

In this paper, we therefore take a different approach and study the
full nonlinear Schr\"odinger--Poisson system of equations
(\ref{spsystem1}--\ref{spsystem2}).  This general type of system has
been studied before, but usually in the context of quantum mechanics,
and always with $\Lambda=0$. Spiegel (1980) was chiefly interested in
discussing its correspondence to the fluid equations.  Its convergence
limits, well posedness, and semiclassical limits have been studied by
Jungel, Mariani \& Rial (2001), Jungel \& Wang (2002) and Li and Lin
(2003). Similar properties of simpler linear Schr\"odinger--Poisson
systems have also been studied by Castella (1991) and by Abdallah,
Mehats \& Pinaud (2005).  The nature of the coupling between the
equations (\ref{spsystem1}) and (\ref{spsystem2}) means that the
\schro equation has what is effectively a time-dependent potential;
solution methods for these have been studied by e.g. Park (2001) and
Devoto \& Pomorisac (1992).  More recently, Erhardt and Zisowsky
(2006) have explored numerical solution techniques for a
spherically-symmetric Schr\"odinger--Poisson system describing the
time evolution of an electron in a classical polar crystal.

Adopting our more direct approach leads inevitably to more
mathematical complication, but we believe this is more than offset by
a clearer physical interpretation. In particular, in our approach the
value of the parameter $\nu$ is unimportant, since it does not appear
in our analytic solutions for the wavefunction $\psi$, which is
contrary to previous approaches. Also contrary to previous authors, we
will not recast our equations in comoving coordinates, since we feel
that this allows for a more straightforward physical
interpretation of our results.  Finally, it is important to stress
that, in our cosmological context, the coupled Schr\"odinger--Poisson
equations (\ref{spsystem1}--\ref{spsystem2}) describe a fully
classical system. Although we wish to utilise any possible links
between quantum mechanics and the Schrodinger description of
cosmological structure formation, caution must be used when drawing
parallels with familiar quantum mechanical ideas, such as the
uncertainty principle.

\section{Cosmological solutions}\label{sec:cosmosolns}

In this section, we discuss the solutions to the
Schr\"odinger--Poisson (SP) system (\ref{spsystem1}--\ref{spsystem2})
in the simple case of a homogeneous and isotropic cosmological dark
matter fluid. In this case, the $P$ term in the SP system is in fact
identically zero, since $|\psi|^2=\alpha^2=\rho$, as a physical
observable, must also be homogeneous. Hence the wave-equation becomes
linear without any need for approximations.  On the other hand, the
velocity potential $\phi$ and the gravitational potential $V$ are not
physical observables, and so may be non-homogeneous; we may only
assume that they are functions of time $t$ and radial distance $r$
only. Thus, we take the ansatz $\alpha = \alpha(t)$, $\phi=\phi(t,r)$
in the wavefunction $\psi$ and $V=V(t,r)$. Consequently the
Laplacian operator in the SP system is simply
\begin{equation}
\nabla^2 = \frac{\partial^2}{\partial r^2} + \frac{2}{r}\frac{\partial
}{\partial r}.
\end{equation}
Although the nonlinear quantum pressure term drops out
of the SP system, the system itself remains nonlinear due to the
coupling of the gravitational potential $V$ to the wavefunction which
is provided by the Poisson equation.
As mentioned above, we will work in `real' (i.e. non-comoving)
coordinates.

Ideally, we would like to obtain results for the most general case,
i.e. that of a universe with arbitrary spatial curvature and a
non-zero cosmological constant, from which all other cases can be
derived. It is well-known, however, that such a solution would include
elliptic functions, with which it is difficult to work analytically.
Consequently, we will not consider the open universe case. Indeed, to
maintain algebraic simplicity, we consider only a spatially-flat
universe with a non-zero cosmological constant and a closed universe
with zero cosmological constant.

\subsection{Spatially-flat universe}

Rather than solving the nonlinear SP system directly, in this case we
can obtain the required solution very quickly by simply assuming the
well-known cosmological result that the fluid density in this case
evolves with (a globally defined) cosmic time $t$ as
\begin{equation}
\label{modifiedpoisson}
\rho(t) \equiv |\psi|^2 = \frac{\Lambda c^{2}}{8 \pi G} {\rm cosech}^{2}
\left(\frac{3}{2}\sqrt{\frac{\Lambda c^{2}}{3}}t
\right).
\end{equation}
This can then be substituted into the modified Poisson equation
(\ref{spsystem2}) to obtain an equation for the gravitational potential
$V$. The solution of this equation is easily found and can then be
substituted into the (in this case linear) wave-equation
(\ref{spsystem1}) to obtain an equation for the velocity potential
$\phi$, which is also easily solved. The resulting
wavefunction and gravitational potential are found to be
\begin{eqnarray}
\label{flatLambdasolution}
  \psi &= & \sqrt{\frac{\Lambda c^{2}}{8 \pi G} }{\rm cosech} (\lambda
  t) \exp \left[\frac{i}{\nu}\sqrt{\frac{\Lambda c^{2}}{12}} {\rm
      coth}(\lambda t)r^{2} \right], \\ V & =
  &\frac{\Lambda c^2}{12}[{\rm coth}^{2}(\lambda
    t)-3]r^{2}, \notag
\end{eqnarray}
where $\lambda \equiv \frac{3}{2}\sqrt{\Lambda c^2/3}$.

The solutions in the special case of a spatially-flat universe with
zero cosmological constant are easily obtained from these results by
performing a series expansion of the functions $\rho = \alpha^2$ and
$\phi/r^2$ around $\lambda=0$, which read
\begin{eqnarray}
 \alpha^2 &=& \frac{1}{6 \pi G t^2} - \frac{c^2 \Lambda}{24 \pi G} + O\left(\Lambda^{3/2}\right), \\
\frac{\phi}{\it{r^2}} &=& \frac{1}{3 t} + \frac{c^2 t \Lambda}{12}+  O\left(\Lambda^2 \right). \notag
\end{eqnarray}
Setting $\Lambda$ to zero, we thus obtain the 
following solution to the SP system:
\begin{eqnarray}
\label{flatsolution1}
\psi &=&\frac{1}{\sqrt{6 \pi G t }}\exp \left( \frac{i r^2}{3 t \nu} \right), \\
\label{flatsolution2}
V &=&\frac{r^2}{9 t^2}.
\end{eqnarray}
This result is easily checked by deriving it afresh by assuming the
well-known result $\rho \propto t^{-2}$ in a spatially-flat universe
with $\Lambda=0$ and following an analogous procedure to that used
above.  In any case, we note that, at at fixed $t$, the potential $V$
has the form of a harmonic oscillator potential in $r$.  Also of
interest is the fact that the wavefunction itself is similar to the
Green's function for the diffusion equation.

\subsection{Closed universe}

To obtain solutions for a closed universe, it is first useful to make
a change of coordinate from cosmic time $t$ to a `conformal time'. In the
usual cosmological approach, the dimensionless conformal time is
defined as
\begin{equation}
\label{usualconformal}
\eta=\int\frac{c\,\mathrm{d}t}{R},
\end{equation}
where $R$ is the scale factor for the closed universe and $c$ is the
speed of light. In the Newtonian context in which we are working,
however, $c$ itself has no special meaning, and so we need to define a
more appropriate conformal time variable. Introducing a constant, $a$,
with dimensions of velocity, we can define a new dimensionless conformal
time $\eta$ as
\begin{equation}
\label{conformaltime}
\eta= a \left( \frac{3M}{4\pi} \right)^{-1/3}\int\alpha^{2/3}\mathrm{d}t,
\end{equation}
where the constant $M \equiv (4\pi/3)\rho R^3$. For notational
brevity, we also introduce the constant $\mu \equiv 3M/(4\pi)$.

In an analogous manner to our analysis of the spatially-flat case, rather
than solving the SP system directly, we begin by assuming the
well-known form of the density evolution in a closed FRW model, as a
function of conformal time $\eta$, namely
\begin{equation}
\label{closedalpha}
\rho\equiv \alpha^2 = \frac{A^2}{\sin^6\!\left(\eta/2\right)},
\end{equation}
where $A$ is a constant. Following the same procedure as that used in
the spatially-flat case, one finds that the wavefunction and
gravitational potential that solve the SP system are
\begin{eqnarray}
\psi & = & \frac{A}{\sin^3(\eta/2)}\exp\left[
\frac{i}{\nu}\frac{a\mu^{-1/3}A^{2/3}\cos(\eta/2)}{2\sin^3(\eta/2)}
r^2\right],\label{psisolnc1}\\
V & = & \frac{a^2\mu^{-2/3}A^{4/3}}{4\sin^6(\eta/2)}r^2,
\label{vsolnc1}
\end{eqnarray}
together with the following condition on the constants:
\begin{equation}
\frac{a}{(\mu A)^{1/3}} = \sqrt{\frac{8\pi G}{3}}.
\label{constants}
\end{equation}
Indeed, one can recover the spatially-flat case (with zero
cosmological constant) by taking the limits $a\to 0$ and $A \to 0$ in
such a way that $A/a^3$ remains finite.

It will be convenient to write the solutions (\ref{psisolnc1}) and
(\ref{vsolnc1}) in a more transparently cosmological form. To assist
in this endeavour, it is first useful to obtain parametric equations
for $R$ and $t$ in terms of the conformal time $\eta$. Using the
expressions (\ref{conformaltime}) and (\ref{closedalpha}), and the
relationship (\ref{constants}) between the constants, one quickly
finds the well-known results
\begin{eqnarray}
R & = & \frac{2GM}{a^2} \sin^2 (\eta/2), \label{revol}\\
t & = & \frac{GM}{a^3} (\eta - \sin\eta), \label{commontime}
\end{eqnarray}
where we have assumed the initial condition that $\eta=0$ when $t=0$.
Indeed, from the first result we see that our velocity parameter
$a=\sqrt{2GM/R_{\rm max}}$, where $R_{\rm max}$ is the maximum value of
$R$ attained during the evolution.

Using the results (\ref{revol}) and (\ref{commontime}), together with
the (\ref{conformaltime}), (\ref{closedalpha}) and (\ref{constants}),
one can easily find the expressions for the Hubble and density
parameters to be
\begin{eqnarray}
H & \equiv & \frac{1}{R}\frac{dR}{dt} = 
\frac{1}{R}\frac{d\eta}{dt}\frac{dR}{d\eta}= \frac{a^3}{2GM}
\frac{\cos(\eta/2)}{\sin^3(\eta/2)}, \\
\Omega & \equiv & \frac{8\pi G\rho}{3H^2} = \frac{1}{\cos^2(\eta/2)}.
\end{eqnarray}
Both parameters have the correct limits at the `big bang', $\eta
=0$, and at `turnaround', or the maximal expansion point, $\eta =
\pi$.

Using these results, it is straightforward now to express the 
wavefunction (\ref{psisolnc1}) and gravitational potential
(\ref{vsolnc1}) in terms of standard cosmological variables as
\begin{eqnarray}
\label{closedsolution1}
\psi &=&\sqrt{\frac{3\Omega H^2}{8\pi G}} \exp \left( \frac{i Hr^2}{2 \nu} \right), \\
\label{closedsolution2}
V &=&\tfrac{1}{4}\Omega H^2 r^2.
\end{eqnarray}
It is simple to verify that these results reduce to the
flat-space solutions (\ref{flatsolution1}) and (\ref{flatsolution2})
in the case where $\Omega=1$ and $H=2/(3t)$.

\subsection{General case in terms of cosmological parameters}

It is, in fact, possible to obtain expressions for $\phi$ and $V$, in
terms of standard cosmological parameters, that are valid for any
pressureless cosmological fluid.

For $\phi$, we have the obvious result
\begin{equation}
\phi= \half H r^2,
\end{equation}
which follows since $\bgrad \phi$ gives the velocity, which should be
$H\br$ in the cosmological case. For $V$, we can use the Bernoulli
equation (\ref{bernoulli}). Using the above expression for $\phi$, we
can immediately deduce
\begin{equation}
V=-\half r^2 \left(\dot{H}+H^2\right),
\end{equation}
Since $H\equiv \dot{R}/R$, we have
\begin{equation}
\dot{H}+H^2=\frac{\ddot{R}}{R}=-qH^2,
\end{equation}
where $q=-\ddot{R}R/\dot{R}^2$ is the usual deceleration parameter.
Thus, the expression for $V$ becomes
\begin{equation}
V=\half \, q H^2 r^2,
\end{equation}
which can be seen to agree with all the individual expressions given
above, and is also valid in the $\Lambda \neq 0$ case, for arbitrary
spatial curvature. In that case,
\begin{equation}
q=(\tfrac{1}{2}\Omega_m-\Omega_{\Lambda})
\end{equation}
where $\Omega_m=8\pi G\rho/(3H^2)$ is the matter density, and
$\Omega_{\Lambda}=\Lambda c^2/(3H^2)$.

In the same spirit of expressing results in terms of cosmological
parameters, the general result for $\alpha$ is
\begin{equation}
\alpha = {\rm constant} \times \exp\left(-\frac{3}{2}\int H dt\right),
\end{equation}
although this is less explicit than the results for $V$ and
$\phi$. The overall wavefunction in the general case is thus
\begin{equation}
\psi 
= {\rm constant} 
\times \exp\left(-\frac{3}{2}\int H dt+\frac{i}{2\nu}Hr^2 \right).
\end{equation}

\section{Modelling a spherical overdensity}\label{sec:sphericalmodel}

We now use the two cosmological solutions found in
Section~\ref{sec:cosmosolns} as the basis for constructing an
analytical wave-mechanical description of the so called `top-hat'
spherical overdensity model, commonly used to study nonlinear
gravitational collapse (e.g. Peebles 1980; Padmanabhan 1993).  The
reader should keep in mind, however, that in all that follows we are
simply describing ordinary Newtonian fluid evolution, albeit in a
novel way.

The model system comprises: an infinite
outer fluid region of homogenous density; an inner fluid sphere, also
of homogenous density, but overdense relative to the outer region; and
a spherical shell of vacuum which separates the two.fluids. The inner fluid
sphere is usually `compensated', that is, we can imagine forming the
inner fluid region by removing the mass from the vacuum (gap) region
and adding it (homogeneously) to the mass within the inner radius of
the gap. As a consequence of Gauss' theorem (or equivalently of Birkhoff's
theorem in general relativity), and because both the inner and outer
fluid regions in this model are homogeneous, each region of fluid
effectively behaves as a distinct `universe', evolving independently of
the other according to cosmological solutions such as those found in
Section~\ref{sec:cosmosolns}.  The inner `island universe' region
remains isolated from the outer universe in which it is embedded via
the compensation condition, which effectively ensures that
the outer radius of the gap region evolves in such a way as to keep
the two fluid regions distinct at all times.

In the cosmological context, the outer region is generally taken to be
a spatially-flat Friedmann universe, with compensation ensuring that
the mean density of the system as a whole is kept at the critical
value. The overdense inner region will evolve as a closed universe.
It is also usual to demand that both the inner and outer universes
have the same origin in time, i.e. that their `big bangs' are
synchronized. This simple spherical overdensity forms the basis of the
`Swiss cheese' model widely used in studies of cosmological structure
formation. For simplicity, we will assume throughout that $\Lambda=0$.

\subsection{Specification of the spherical overdensity model}

\begin{figure}
\begin{center}
\includegraphics[width=60mm]{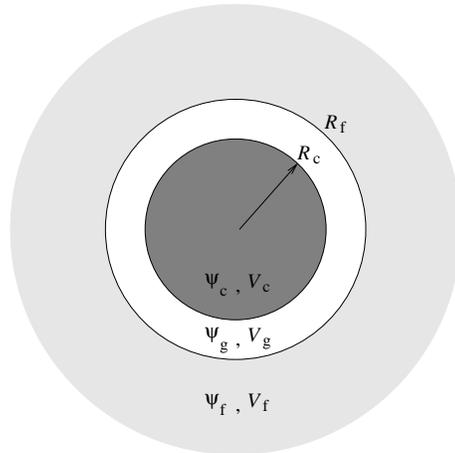}
\caption{Illustration of the spherical overdensity. 
The wavefunction and gravitational potential that solve the SP system
in the inner, vacuum, and outer regions, respectively, are
  $(\psi_{\rm{c}}, V_{\rm{c}})$, $(\psi_{\rm{g}},
  V_{\rm{g}})$, and $(\psi_{\rm{f}}, V_{\rm{f}})$. The radii
  $R_c$ and $R_f$ label the boundaries of the vacuum region.
\label{overdensity}}
\end{center}
\end{figure}

Our aim is to obtain a global wavefunction and gravitational potential
that solve the SP equations (\ref{spsystem1}--\ref{spsystem2}) for
this physical system; we do this by constructing a piecewise solution
from the cosmological solutions derived in
Section~\ref{sec:cosmosolns} and a vacuum solution. An illustration of
the physical system is shown in Fig.~\ref{overdensity},
where $(\psi_{\rm{c}}, V_{\rm{c}})$, $(\psi_{\rm{g}}, V_{\rm{g}})$ and
$(\psi_{\rm{f}}, V_{\rm{f}})$ represent the wave-mechanical solutions
to the Schr\"odinger--Poisson system in the inner (closed), vacuum (gap) and
outer (flat) regions respectively. The radii $R_c$ and $R_f$ label the
inner and outer boundaries of the vacuum region respectively. We
will denote by $(\psi,V)$ the
global wavefunction and gravitational potential 
that we construct, in a piecewise
fashion, from the solutions for each region.

Our first task is to establish a global time coordinate for the
system. This is, in fact, straightforward, since equation
(\ref{commontime}) already links the `ordinary' (cosmic) time
coordinate $t$ appearing in the solution for the spatially-flat outer
region with the conformal time $\eta$ in the solution for the closed
inner region.  This condition establishes the correct phasing between
the two regions by setting the conformal time equal to zero at the
zero of ordinary time. Since the overdense inner region is compensated
precisely by the surrounding vacuum region, our system evolves
correctly in that the volume of the vacuum region vanishes as $t \to
0$ (or $\eta \to 0$).  This is easily verified as follows. The density
of the flat universe varies as $\rho=1/\sqrt{6 \pi G t^2}$, and so
\begin{equation}
R_f=\left(\frac{9GM}{2}\right)^{1/3}t^{1/3},
\end{equation}
where $M=(4\pi/3)\rho_cR_c^3$ is the total mass of the inner fluid
region.  This expression for $R_f$ can be compared to the equation
(\ref{revol}) for the evolution of outer radius, $R_c$, of the closed
inner region. One finds that $R_f/R_c$ is a function of conformal time
$\eta$ only, and for small $\eta$ is given by
\begin{equation}
\frac{R_f}{R_c}=1+\frac{\eta^2}{20}+\cdots,
\end{equation}
which correctly tends to unity as $\eta \to 0$, i.e. as the
big bang is approached.

In the context of the finite inner fluid region of total mass $M$
evolving as an isolated closed universe, it is worth noting that the
role of the velocity parameter $a=\sqrt{GM/R_{\rm max}}$, introduced
in (\ref{conformaltime}), now becomes clear: it is the Newtonian
escape velocity at the surface of the inner fluid when it reaches its
maximum radius. The limit in which the inner fluid region 
becomes spatially-flat corresponds to the escape velocity $a$ at this
point tending to zero. (We also note that $a=c$ would correspond to
the maximum radius $R_{\rm max}$ being the Schwarzschild radius
for a mass $M$).

\subsection{Boundary conditions}

To construct piecewise global solutions for the wavefunction $\psi$
and gravitational potential $V$ describing the full system, it is
necessary to determine the appropriate boundary conditions on these
functions at the two moving surfaces $R_c$ and $R_f$.

We first consider the boundary conditions on the gravitational
potential. At each boundary in this highly idealized system, the fluid
`boundary layer' is infinitessimally thin, and effectively
massless. Consequently, it cannot support a force acting on it, which
in turn implies that $\partial V/\partial r$, and hence also $V$
itself, must be continuous at each boundary. The second derivative of
$V$ may, however, have a discontinuity, which will correspond to the
gradient of the force being discontinuous at the boundary. This is
indeed physically correct, since in the vacuum region there will be
tidal forces, but clearly none in the homogeneous regions.  The
gravitational potential $V$ is determined by the Poisson equation
(\ref{spsystem2}), with $\Lambda=0$, which for our spherical geometry
becomes
\begin{equation}
\frac{\partial^2V}{\partial r^2}
+\frac{2}{r}\frac{\partial V}{\partial r}=4\pi G\alpha^2.
\end{equation}
A discontinuity in the second derivative of $V$ thus implies
that $\alpha$ should also possess a discontinuity at each boundary, as
expected.

We turn now to the boundary conditions on the wavefunction $\psi$.
Griffiths (1992) has previously considered the boundary conditions on
a time-independent wavefunction in the presence of a potential
possessing a discontinuity (modelled as a Heaviside function, i.e. the
derivative of a Dirac delta function).  In our time dependent case,
however, Griffiths' approach will not yield the appropriate boundary
conditions. To determine the consequences of the discontinuity in square-root
density $\alpha$ at each boundary, we must turn to the equation
governing the time development of $\alpha$, which is
\begin{equation}
\label{alphaeqn}
\frac{\partial\alpha}{\partial t}=-\frac{\partial\alpha}{\partial
  r}\frac{\partial\phi}{\partial
  r}-\frac{\alpha}{r}\frac{\partial\phi}{\partial
  r}-\frac{\alpha}{2}\frac{\partial^2\phi}{\partial r^2}.
\end{equation}
For our solution $(\psi, V)$ for the full system to be consistent, the
form for $\alpha$ that we know to be correct at the boundaries must
satisfy this equation at all times. Consider, for example, the
boundary $r=R_f(t)$, where we make a transition from vacuum to
homogeneous fluid. The form of $\alpha$, as $r$ increases, is
\begin{equation}
\label{alphaform}
\alpha\!\left(t,r\right)=f\!\left(t\right)H\!\left(r-R_f\!\left(t\right)\right)
\end{equation}
where $f(t)$ describes the time evolution of $\alpha$ within the outer
(flat) region of homogeneous fluid and $H$ is the Heaviside step function.
Substituting this form into (\ref{alphaeqn}), one finds
\begin{equation}
\vspace*{-0.2cm}
\hspace*{-3cm}\left(\frac{\mathrm{d}f}{\mathrm{d}t}
+\frac{f}{r}\frac{\partial\phi}{\partial r}
+\frac{f}{2}\frac{\partial^2\phi}{\partial r^2}\right)
H\!\left(r-R_f\!\left(t\right)\right)
\nonumber
\end{equation}
\begin{equation}
\hspace{1.75cm}+\frac{f}{r}\left(f\frac{\partial\phi}{\partial r}
-R_f\frac{\mathrm{d}R_f}{\mathrm{d}t}\right)\delta\!
\left(r-R_f\!\left(t\right)\right)=0.
\end{equation}
In order for the postulated form (\ref{alphaform}) of $\alpha$
to be consistent, the coefficients of the step and delta functions need to
vanish separately. This follows immediately for the coefficient of $H$,
since on closer inspection this is seen to be the equation for the
time development of $\alpha$ in the homogenous region, where
$\alpha=f$. The coefficient of the delta function needs to be
evaluated at the moving boundary $r=R_f(t)$, which means that one
requires $dR_f/dt = d \phi/dr$ there. This is, however, exactly the
statement that the boundary is moving at the speed which is given by
the velocity potential at that point, and so this coefficient vanishes
also. A similar analysis shows that this form for $\alpha$ is also
acceptable at the other boundary.
 
It now remains only to consider what, if any, conditions must be
satisfied by the phase of the wavefunction, the velocity
potential $\phi$, at the boundaries. The equation determining the
evolution of $\phi$ is simply the Bernoulli equation (\ref{bernoulli}),
which in this case reads
\begin{equation}
\label{velocitypotentialwithtime}
\frac{\partial\phi}{\partial t}=-\tfrac{1}{2}\left(\frac{\partial\phi}
{\partial r}\right)^2-V.
\end{equation}
Examining this equation, it becomes clear first of all that it is not
necessary to consider the possibility of singularities in $\phi$ at
the boundaries: since $V$ has continuous zeroth and first derivatives,
then $\phi$ will also remain singularity free. Additionally, in
regions where $\psi=0$, i.e. the vacuum region, $\phi$ may take
any value at all, since it is not defined there.
 
To summarize, we have shown that the only physical boundary conditions
which must be satisfied for the solution $\psi, V$ for this system are
those of continuity of the gravitational potential $V$ and its first
derivative. The wavefunction $\psi$ itself need not be continuous at
the boundaries. We have also established that our wave mechanical
description of the system is thus far self consistent, by
demonstrating that a step discontinuity in the modulus of the
wavefunction at both boundaries satisfies the time dependent
Schr\"odinger--Poisson system, as expected.

\subsection{Forming the piecewise solution}

Having established the boundary conditions on the wave mechanical
description $(\psi, V)$ of the spherical overdensity system, we can
now consider how we can form this wavefunction from those which
separately describe each region.  

We first consider the wavefunction $\psi$. The modulus of the
wavefunction, the root-density $\alpha$, is already specified
everywhere, and thus we have no further work to do here. The phase of
the wavefunction, the velocity potential $\phi$, is fully specified in
the inner and outer fluid regions by (\ref{closedsolution1}) and
(\ref{flatsolution1}) respectively, but it is undefined in the vacuum
region since there is no fluid there. Moreover, our analysis above has
shown that we do not require continuity of phase of the wavefunction
at either boundary. 

We turn now to the gravitational potential $V$. We first note that, since
$\alpha=0$ in the vacuum region, the gravitational
potential in this region must satisfy the Laplace equation, so that
\begin{equation}
\label{Vsolution}
V_{\mathrm{g}}=C+\frac{B}{r},
\end{equation}
where $B$ and $C$ are functions of time only and must be fixed by
appropriate boundary conditions.  We showed above that, for a
physically meaningful solution, we require $V$ and $\partial
V/\partial r$ to be continuous at each boundary. Consequently, we must
match the vacuum solution (\ref{Vsolution}) and its radial derivative at
each boundary with the corresponding solution for the inner or outer fluid
region. Matching at the inner boundary yields the vacuum solution
\begin{equation}
\label{innerV}
V_{\mathrm{g,1}}=GM\left(\frac{3}{2R_c}-\frac{1}{r}\right),
\end{equation}
whereas matching at the outer boundary gives
\begin{equation}
V_{\mathrm{g,2}}=GM\left(\frac{3}{2R_f}-\frac{1}{r}\right),
\end{equation}
where $R_c$ and $R_f$ are functions of time. These two results have
the same functional form inside the vacuum, but differ by a
constant. This is problematic: since $V$ is defined through the
vacuum, these solutions, unlike those for $\phi$, do need to match.
This problem is resolved, however, by appealing to the extra degrees
of freedom offered to us by the fact that the velocity potential in
each fluid region is fixed only up to the addition of an arbitrary
function of time.  From equation (\ref{velocitypotentialwithtime}) for
the time evolution of $\phi$, we see that for the vacuum solutions for
$V$ derived at each end of the boundary to match, we must simply
adjust one or both of the velocity potentials $\phi$ at the two ends
of the vacuum by functions of time alone such that their difference is
altered by the following
\begin{equation}
\label{gaugechange}
\Delta\phi=\int_0^t\left(\frac{3}{2R_c}-\frac{3}{2R_f}\right)\mathrm{d}t.
\end{equation}
Physically, at the level of the \schro equation, this corresponds to a
gauge change, which alters the potential by a function of time, by
introducing a phase term (the integral of the change in time) into the
wavefunction. This phase term is introduced relative to the
cosmological solution wavefunction in the relevant region.

The solutions for individual regions can now be put together to form a
global solution with all boundary conditions satisfied. The solution
(\ref{closedsolution1}--\ref{closedsolution2}) is taken as the global
solution $(\psi, V)$ for $r\leq R_c $. In the vacuum region, between
$R_c$ and $R_f$, the velocity potential is undefined, and we match $V$
appropriately at the inner boundary so that the gravitational
potential (\ref{innerV}) holds throughout the gap. We then employ the
gauge freedom in $\phi$ to adjust the velocity potential in the outer
region, $r \geq R_f$, so that the global gravitational potential is
continuous. We do this by subtracting (\ref{gaugechange}) from the
phase of the wavefunction in (\ref{flatsolution1}), which then becomes
the solution for $\phi$ in the outer region. Equation
(\ref{velocitypotentialwithtime}) can then be used to find the
corresponding gravitational potential in the outer region, which is
correctly matched at the boundaries.  The resulting global solution
for $\psi=\alpha e^{i \phi / \nu}$ and $V$, defined piecewise in
$\alpha, \phi$ and $V$ for $r \leq R_c$, $R_{c} < r < R_{f}$ and $r
\ge R_f$ respectively, is thus given by
\begin{eqnarray}
	\label{alphapiecewise}
	\alpha& =& \left\{ \begin{array}{l}
\sqrt{\frac{3\Omega H_c^2}{8\pi G}}, \\[2mm]
	0, \\[2mm]
\sqrt{\frac{3H_f^2}{8\pi G}}, 
	\end{array} \right.\\[2mm]
	\label{phipiecewise}
	\phi& =& \left\{ \begin{array}{l}
	\frac{1}{2}H_cr^2, \\[2mm]
	\mbox{undefined}, \\[2mm]
	\frac{1}{2}H_fr^2 + \frac{3GM}{2}\int_0^t 
\left(\frac{1}{R_f}-\frac{1}{R_c}\right)\,dt',
	\end{array} \right.\\[2mm]	
	\label{Vpiecewise}
	V& =& \left\{ \begin{array}{l}
	GM\frac{r^2}{2R_c^3}, \\[2mm]
	GM\left( \frac{3}{2R_c} - \frac{1}{r} \right), \\[2mm]
	\frac{1}{2}GM\left(\frac{r^2}{R_f^3} + \frac{3}{2R_c} - \frac{3}{2R_f}
        \right). 
	\end{array} \right.	
\end{eqnarray}

It is worth noting the physically interesting result that the
`correction term' in the velocity potential $\phi$ in the outer region
$r \ge R_f$ is proportional to the difference in the conformal time
variables in the inner and outer regions. Also of interest is that,
for similar physical systems, it is clear that we will not require any
particular conditions on the wavefunction at the boundaries. As
illustrated by our (somewhat pathological) example, step
discontinuities ensure that the wavefunction and its first derivative
certainly are not required to be continuous there, although we can
imagine situations where they may be; an example of this might be a
smoothly varying density of fluid with a boundary represented by an
abrupt change in fluid velocity at a given radius. This behaviour is
in marked contrast to familiar systems in quantum mechanics, where it
is the gravitational (or other) potential which generally is
considered to have discontinuities, but the wavefunction must
satisfy physical boundary conditions.

\subsection{Comparison with quantum \schro systems}

It is interesting to consider in more detail how our results compare
with what may be considered a similar system in standard quantum
mechanics. Surprisingly, the problem of which boundary conditions to
apply at a moving boundary is a matter of debate in the literature: an
example of the confusion is provided by Samura and Ohmukai (2006). The
authors consider reflection and transmition of an incident
wavefunction at a moving potential step using the standard
time-dependent \schro equation.

\begin{figure}
	\label{Samuraplot}
\begin{center}
\includegraphics[width=60mm]{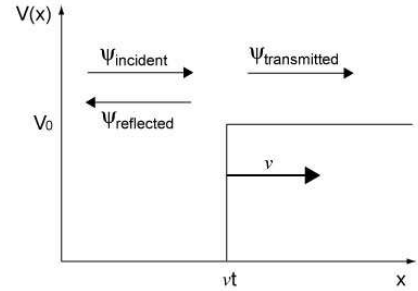}
\caption{Reflection and transmission of a wavefunction at a boundary
  of height $V_{0}$, moving at speed $v$, after Samura and Ohmukai
  (2006). 
}
\end{center}
\end{figure}

They claim that the boundary conditions one would expect to apply,
namely that the wavefunction and its first derivative are continuous,
are not in fact valid. Since the position of the potential step is a
function of $t$, they claim that one cannot use the continuity of the
first derivative because in taking this condition, the boundary is
fixed in time. Instead, they argue, one should use an alternative
boundary condition, that the phase of the wavefunction is continuous
at the boundary, and they proceed to use this to derive expressions
for the reflected and transmitted waves.

Although this condition yields the correct final results, it is
straightforward to show that the authors are incorrect in claiming
that the continuity of the first derivative is not a valid boundary
condition to apply. One can in fact use the standard boundary
conditions to derive the correct results, simply applying a Galilean
transformation to the stationary barrier case.

In this approach, we take the usual \schro equations left and right of
the boundary, but take the three wavefunctions, $\psi_{\mathrm{i}},
\psi_{\mathrm{r}},$ and $\psi_{\mathrm{t}}$ as functions of $x'=
x-vt$. We examine the case in which, if the step were stationary,
$\psi_{\mathrm{t}}$ would be purely evanescent, so that we can take
the limit to an infinite step, which is the case most closely
resembling our cosmological \schro system. By applying the usual
boundary conditions of continuity of the wavefunction and its first
derivative at the boundary, $x=vt$, one finds the following results
for the wavenumbers $k_{\mathrm{r}}$ and $k_{\mathrm{t}}$ in terms
of $k_{\mathrm{i}}$ (setting $\hbar=1$):
\begin{eqnarray}
\label{k2}
k_{\rm r} & = & k_{\rm i} - 2mv, \\
\label{k3}
k_{\rm t} & = & -imv 
+ \sqrt{-m^{2} v^{2} + 2vmk_{\rm i} - {k_{\rm i}}^2 + 2V_{0}m}.
\end{eqnarray}
These yield wavefunctions which satisfy the \schro equations in each
region, and they reduce to the correct results in the case where
$v=0$. A series expansion of the first derivative of the wavefunction
in the case that $V_{0} \rightarrow \infty$ shows that there is a finite
slope at the boundary, demonstrating that this
method works for both the finite and infinite potential step.

This ambiguity in the published literature highlights the fact that
determining which boundary conditions to apply to \schro
wavefunctions, even in the usual quantum mechanical context, is not a
trivial matter. It is interesting that in the case of our classical
gravitational system, neither the matter wavefunction nor its
derivative need be continuous, in contrast to the usual quantum
case. Instead, we require that conditions be placed on the
gravitational potential which is coupled to the wavefunction. 

\section{Multiple fluids}\label{sec:shellcrossing}

The fluid description of cosmological dark matter breaks down
when multistreaming occurs, most notably at shell-crossing.  In
general, the onset of shell-crossing occurs after the onset of
nonlinearity; shell-crossing thus represents a natural `barrier'
beyond which fluid-based methods cannot follow the evolution of structure
into the fully nonlinear regime. Attempts to describe structure in the
quasi-linear regime using a fluid description therefore resort to
approximate methods to extend the approach beyond shell-crossing,
e.g. the Zeldovich, adhesion, and free-particle approximation methods.

The \schro formalism is also based on a fluid description, and
consequently shares these limitations.  Nonetheless, as in other
fluid-based approaches, one can model more general physical systems by
considering them to be made up of multiple fluids. One still cannot
accommodate multi-streaming within each individual fluid, but one can
model some multi-streaming situations by considering each stream as a
separate fluid.  In this section, we therefore extend our previous
description by moving from a single fluid to the multiple fluid case,
inspired by the application of techniques from multiparticle quantum
mechanics.  We do this by assigning a distinct wavefunction to
distinct regions of the CDM fluid, each described by its own
wave-equation, which interact collectively via a common coupled
Poisson equation.

The toy cosmological system illustrated in Fig.~\ref{overdensity} can
always be successfully described using the \schro formulation
described in the previous section because the inner and outer regions
of fluid never overlap, i.e. its inner and outer `shells' never
cross. There is thus a single unique fluid velocity for each point in
the system, throughout its evolution. This formulation would break
down at multistreaming because the Madelung transformation ansatz
(\ref{wavefunctiondefinition}) for the form of the fluid wavefunction
assumes a single-valued velocity field.

A good toy model for illustrating multi-streaming in cosmology can be
constructed, however, by considering a physical system similar to that
discussed in Section~\ref{sec:sphericalmodel}, but with appropriate
initial and boundary conditions to ensure that the two fluid regions
do overlap at some stage. We now describe how to construct a new,
two-fluid description for such a system that accommodates
this multi-streaming.

\subsection{Multiparticle wavefunctions and multiple fluids}

Since we have previously adopted a wavefunction approach to modelling
a single fluid, it is natural to consider whether
multiparticle methods in quantum mechanics can assist in providing a
wavefunction approach to multiple fluids. For simplicity, we
will confine our attention to just two fluids.

Several questions immediately arise, such as what is the correct form
of the quantum pressure and potential terms that should be
used in the `two-particle' case, and whether `entangled' states have
any role to play in the classical evolution of two coupled fluids, or
whether factorisable states are the only ones relevant.

In the single-fluid case, the approach we have been using is strongly
related to the de Broglie--Bohm formulation of quantum mechanics (for a
full account see, for example, Holland 1989).  It is natural,
therefore, to turn first to the many-particle version of de
Broglie--Bohm theory. We now wish to adapt this
approach to obtain an equation that describes the evolution of two
fluids, i.e. describes their velocities and particle densities in time
and space, using a \schro formalism. Of course, we also need to show
how one can obtain individual densities and velocities for each of the
fluids from the resulting equation.

In the two-fluid case, the overall wavefunction  now has the form
\begin{equation}
\psi = \psi(\bmath{r}_1,\bmath{r}_2,t),
\end{equation}
where $\bmath{r}_1$ and $\bmath{r}_2$ are spatial positions for an
element within fluid 1 or fluid 2 respectively, and $t$ is the overall
common time variable; we have thus introduced a $(6+1)$-dimensional
parameter space. As for the single fluid case, we spilt $\psi$ into a
magnitude and phase written as
\begin{equation}
	\label{eqn:amp-phase-rep}
\psi(\bmath{r}_1,\bmath{r}_2,t) = \alpha(\bmath{r}_1,\bmath{r}_2,t)
\exp[i\phi(\bmath{r}_1,\bmath{r}_2,t)/\nu].
\end{equation}
We also introduce the gradient operators $\bgrad_i$ $(i=1,2)$ for each
fluid, where we adopt the convention that bold quantities `live' in
the individual fluid spaces, and non-bold quantities refer to the
total space (with 6 `spatial' dimensions). Using Cartesian coordinates
$\bmath{r}_i=(x_i,y_i,z_i)$ in each fluid space, for example, the
corresponding Laplacian operators (for $i=1,2$) are
\begin{equation}
\bgrad_i^2 =  \frac{\partial^2}{\partial x_i^2}
+ \frac{\partial^2}{\partial y_i^2}
+ \frac{\partial^2}{\partial z_i^2}
\end{equation}

We now assert that the overall two-fluid wave-equation is
\begin{equation}
i \nu \frac{\partial \psi}{\partial t}=
 -\frac{\nu^2}{2} \left( \bgrad_1^2 + \bgrad_2^2\right) \psi
+2V_{1,2} \psi
+P\psi = 0, \label{eqn:2-part-S}
\end{equation}
where $V_{1,2}$ is the common gravitational potential and
the quantum pressure term has the form
\begin{equation}
P = \frac{\left(\bgrad_1^2 + \bgrad_2^2\right) |\psi|} {2 |\psi|}.
\end{equation}
This expression for $P$ is certainly the natural generalisation to the
two-fluid case of the equivalent $P$-term for a single fluid.  It also
agrees with the `many-body quantum potential' $Q$ given in equation
(7.1.4) of Holland (1989), although our quantum pressure term is the
negative of this, since here we are seeking to remove the term that
would otherwise appear in the fluid equations of motion.

A useful quantity relating fluid density and velocity in de
Broglie--Bohm theory is the probability current. The natural
generalisation of the probability current to two fluids, is
\begin{equation}
J = \frac{1}{2mi}\left[ \left( \grad \psi \right)\psi^* -\psi\left(\grad \psi\right)^*\right],
\end{equation}
where $\grad = \bgrad_1 + \bgrad_2$. The interpretation of $J$ as a
current in six dimensions, corresponding to a flow of `probability'
into and out of regions, just as for the single-fluid case, follows
from
\begin{equation}
\frac{\partial |\psi|^2}{\partial t} + \grad \cdot J = 0,
\label{eqn:2-part-cons}
\end{equation}
which is easy to prove by using the two-fluid wave-equation
(\ref{eqn:2-part-S}).

Using the split (\ref{eqn:amp-phase-rep}) into amplitude and phase
terms, we find
\begin{equation}
J =  \alpha^2 \left( \bgrad_1 \phi + \bgrad_2 \phi \right),
\end{equation}
and then identifying the individual
space velocities
\begin{equation}
\bv_1 =  \bgrad_1 \phi, \quad \bv_2 =  \bgrad_2 \phi \label{eqn:v-defs}
\end{equation}
we have
\begin{equation}
J = \alpha^2 \left( \bv_1 + \bv_2 \right) \label{eqn:simple-6d-cont}.
\end{equation}
The conservation equation (\ref{eqn:2-part-cons}) can now be written
\begin{equation}
\frac{\partial \alpha}{\partial t} + \bgrad_1 \dt \left(\alpha^2 \bv_1\right) + \bgrad_2 \dt \left(\alpha^2 \bv_2\right) = 0.
\end{equation}

The other equation we require is that relating the common
gravitational potential $V_{1,2}$ to the densities of the two fluids
sources. The term $V_{1,2}$ (for which we effectively include a copy
for each fluid in the wave-equation) is taken to satisfy
the (single-fluid) equation
\begin{equation}
\bgrad^2 V_{1,2} = 4 \pi G \left( \rho_1 + \rho_2 \right), \label{eqn:potential}
\end{equation}
i.e.\ the potential under which both fluids move is generated by the
sum of their densities.

The wave-equation (\ref{eqn:2-part-S}) and the Poisson equation
(\ref{eqn:potential}) give what seems a plausible `two-particle' quantum
description of two interacting classical fluids. The most basic test
that any many-body equation must satisfy is that it works for a
factored state, for which
\begin{equation}
\psi(\bmath{r}_1,\bmath{r}_2,t)=\psi_1(\bmath{r}_1,t)
\psi_2(\bmath{r}_2,t).
\end{equation}
Inserting this form of $\psi$ into the wave-equation
(\ref{eqn:2-part-S}), 
dividing through by $\psi_1\psi_2$, and applying the usual separation
of variables argument (here to the coordinates $\bmath{r}_1$ and
$\bmath{r}_2$), we obtain the two individual fluid equations
\begin{equation}
\begin{aligned}
\label{eqn:twoindividual}
i \nu \frac{\partial\psi_1}{\partial t} &=
-\frac{\nu^2}{2}\bgrad_1^2 \psi_1
+V_{1,2}\psi_1 +\frac{ \nu^2 \bgrad_1^2 |\psi_1|}{2|\psi_1|} \psi_1,\\
i \nu \frac{\partial\psi_2}{\partial t} &=
-\frac{\nu^2}{2}\bgrad_2^2 \psi_2
+V_{1,2}\psi_2 +\frac{\nu^2 \bgrad_2^2 |\psi_2|}{2|\psi_2|} \psi_2.
\end{aligned}
\end{equation}
The wave-equation (\ref{eqn:2-part-S}) thus makes sense for factored
states. However, generalizing to `entangled states' appears
problematic. This can be seen immediately in equation
(\ref{eqn:potential}) for the gravitational potential, which requires
knowledge of (the sum of) the separate fluid densities $\rho_1$ and
$\rho_2$. Given a general two-fluid wavefunction $\psi$, it is not
clear how these would be obtained. Given $\psi$, we can successfully
form the two separate particle velocities, via the definitions
(\ref{eqn:v-defs}), which only require that the total phase, $\phi$,
be available. However, knowledge of the overall magnitude,
$\alpha^2=|\psi|^2$, is {\em not\/} enough to deduce the separate
densities. Thus, in the general entangled case, we have no way to generate
$\rho_1$ and $\rho_2$ separately from $\psi$, and furthermore, it
appears that we cannot find their sum either (to put into the right
hand side of (\ref{eqn:potential})), so that the equation set becomes
incomplete as it stands.

A possible solution to this problem is to assume continuity in the two spaces
separately, which seems a reasonable physical requirement. Supposing that
\begin{equation}
\frac{\partial \rho_1}{\partial t} + \bgrad \dt (\rho_1 \bv_1) = 0 \quad \text{and} \quad
\frac{\partial \rho_2}{\partial t} + \bgrad \dt (\rho_2 \bv_2) = 0,
\end{equation}
and given initial density distributions $\rho_1$ and $\rho_2$, there
is enough information available to propagate the two densities
individually forward in time, since $\bv_1$ and $\bv_2$ are separately
available. Thus there is {\em in principle\/} an answer available,
although it would not be a useful one numerically.

We note further that the intuitive identification
$\alpha^2=\rho_1\rho_2$ is not even necessarily valid in the
non-factored case, since the 6-dimensional continuity equation
(\ref{eqn:simple-6d-cont}), plus the assumption of continuity for each
fluid separately, can be used to show that
\begin{equation}
\frac{d}{dt}\left[ \ln \frac{\alpha^2}{\rho_1\rho_2}\right] =
-\left(\bv_1\dt\bgrad_1 \ln\rho_2+\bv_2\dt\bgrad_2\ln\rho_1\right), \label{eqn:dot-deriv_result}
\end{equation}
where the $d/dt$ represents a comoving or streamline derivative in the
7-dimensional space:
\begin{equation}
\frac{d}{dt} \equiv \frac{\partial}{\partial t} +
\bv_1\dt\bgrad_1+\bv_2\dt\bgrad_2.
\end{equation}

In the factored case, the right-hand side of
(\ref{eqn:dot-deriv_result}) will vanish, meaning that the
identification of $\alpha^2$ with $\rho_1\rho_2$ is maintained along
the (joint-fluid) motion. However, in the non-factored case, there is
no guarantee that right-hand side of (\ref{eqn:dot-deriv_result}) will
indeed vanish, and we have a further problem in interpreting the
wavefunction.

To summarize our answers to the questions posed at the beginning of
this section, we now have a good two-fluid wave-equation description
of the motion of two coupled fluids, for which factored states make
sense and return sensible equations in each particle space (with
the coupling still present due to the potential, but not explicitly at
the coordinate level). The notion of non-factorisable states looks
problematic, but given that we are attempting to describe classical,
rather than quantum, fluids, this is perhaps unsurprising.

\subsection{Numerical two-fluid evolution with spherical symmetry}\label{sec:numevol}

We now apply the two-fluid wave-equation formalism developed above to
model explicitly the evolution of a spherically symmetric system
consisting of two physically identical, mutually self-gravitating
fluids, each of which is described by individual wavefunctions
$\psi_1$ and $\psi_{2}$ respectively. As previously, the wavefunctions
are given by
\begin{eqnarray}
\psi_1& = & \alpha_{1}e^{i \phi_{1}/\nu}, \\
\psi_2& = & \alpha_{2} e^{i \phi_{1}/\nu}, 
\end{eqnarray}
where $\alpha_{1}(r,t)$ and $\phi_{1}(r,t)$ and $\alpha_{2}(r,t)$ and
$\phi_{2}(r,t)$ are the square-root densities and velocity potentials
for each fluid. The equations that these wavefunctions satisfy are
given by (\ref{eqn:twoindividual}), which must, in general, be solved
numerically. 

The simplest way to solve the wave-equations numerically is to
take their real and imaginary parts, and to solve for the amplitude
and phase of each wavefunction. We therefore solve the following
system of five coupled equations for the evolution of the two fluids:
\begin{eqnarray}
\label{alpha1}
\frac{\partial\alpha_1}{\partial t}& =&
-\frac{\partial\alpha_1}{\partial r}\frac{\partial\phi_1}{\partial
  r}-\frac{\alpha_1}{2}\frac{\partial^2\phi_1}{\partial
  r^2}-\frac{\alpha_1}{r}\frac{\partial\phi_1}{\partial r}\\
\label{phi1 }
\frac{\partial\phi_1}{\partial t} 
& = & -V-\tfrac{1}{2}\left(\frac{\partial\phi_1}{\partial r}\right)^2 \\
\label{alpha2}
\frac{\partial\alpha_2}{\partial
  t}&=&-\frac{\partial\alpha_2}{\partial
  r}\frac{\partial\phi_2}{\partial
  r}-\frac{\alpha_2}{2}\frac{\partial^2\phi_2}{\partial
  r^2}-\frac{\alpha_2}{r}\frac{\partial\phi_2}{\partial r}\\
\label{phi2}
\frac{\partial\phi_2}{\partial
  t}&=&-V-\tfrac{1}{2}\left(\frac{\partial\phi_2}{\partial
  r}\right)^2\\
\label{v12}
\frac{\partial^2V}{\partial r^2} +\frac{2}{r}\frac{\partial
  V}{\partial r} & = & 4\pi G \left( \alpha_{1} ^{2} + \alpha_{2}^{2}
\right)
\end{eqnarray}
It is worth noting that equation (\ref{v12}) for the gravitational
potential $V$ does not contain a time derivative and hence, for fixed
$t$, it is, in fact, an ordinary differential equation in $r$.
Moreover, we note that the parameter $\nu$ does not appear in the 
above system of equations.

To solve the system of equations (\ref{alpha1}--\ref{v12}) numerically
for the five variables $\alpha_{i}(r,t)$, $\phi_{i}(r,t)$ $(i=1,2)$
and $V(r,t)$, one must first specify the appropriate initial and
boundary conditions on the solution. The initial conditions at some
time $t=0$ (say) are assigned by specifying $\alpha_{i}(r,0)$ and
$\phi_{i}(r,0)$ $(i=1,2)$. At any given $t$ (including $t=0$),
knowledge of the $\alpha_i$ is sufficient to determine the
gravitational potential $V$ by solving (\ref{v12}). In this
spherically-symmetric case, the general solution of (\ref{v12}), at a
given time, is
\begin{equation}
\label{vdint}
V=\int\frac{4\pi G\int\left(\alpha_1^2+\alpha_2^2\right)r^2\mathrm{d}r
  +c_1}{r^2}\mathrm{d}r+c_2,
\end{equation}
where $c_1$ and $c_2$ are arbitrary functions of time. Clearly
$c_2$ is merely an overall additive term which fixes the zero of
gravitational potential, and $c_1$ expresses whether there is a point
mass at the origin. We choose $c_1=c_2=0$ to fix $V$ in a simple
manner. Furthermore, we will assume that the total fluid density is
non-zero at the origin. This means that the dominant behavior of the
inner integrand is $\propto r^2$, and thus that the dominant behavior
of the outer integrand is $\propto r$. With these assumptions, one can
perform the double integration in (\ref{vdint}) using a set of two
interleaved Simpson's rules, which we have found to work to high
accuracy.

It remains to consider the boundary conditions on $\alpha_{i}$ and
$\phi_{i}$ at each end of the spatial domain, i.e. at $r=0$ and
$r=r_{\rm max}$. We first focus on the conditions that must apply at
the origin. There is a physical requirement at $r=0$ that, in the
absence of a singularity there, it should be neither a source nor a
sink. Since, as mentioned above, we are assuming the total fluid
density $\alpha_{1}^{2} + \alpha_{2}^{2} > 0$ at the origin, we thus
require $\partial\phi_1/\partial r = 0$ and $\partial\phi_2/\partial r
= 0$ (i.e. zero fluid velocities) there. There is also an arbitrary
additive constant in each $\phi_i$ to fix and we do this by requiring
that $\phi_i = 0 $ at the origin. These two boundary conditions for
each $\phi_i$ are compatible with the equations of motion, since
$\partial\phi_i/\partial r = 0$ and $V=0$ at $r=0$ combine to mean
that $\partial\phi_i/\partial t$ is also zero there, so $\phi_i = 0$
will be maintained. A further detail that has to be addressed at $r=0$
is how to evolve each $\alpha_i$ forward in time
given that $1/r$ appears in one of the terms of $\partial
\alpha_i/\partial t$. Since $\phi_i$ and $\partial\phi_i/\partial r$
are both zero at $r=0$ we can assume that the dominant behavior of
$\phi_i$ is $\propto r^2$. In this case, we can combine the last two
terms of $\partial\alpha_i/\partial t$ to obtain for each fluid
\begin{equation}
\left.\frac{\partial\alpha_i}{\partial t}\right|_{r=0}
=\left.-\frac{3\alpha_i}{2}\frac{\partial^2\phi_i}{\partial r^2}\right|_{r=0}.
\end{equation}
If $\phi_i$ is in fact proportional to a higher power of $r$, then
this formula is still correct, since it then returns zero. It is worth
noting in the above prescriptions, there is no requirement that the
$\alpha_i$ have zero gradient at $r=0$. Thus a cusp in the $\alpha_i$
distributions at the origin is physically allowed, and does not mean
that there is a singularity there.

We now consider the boundary conditions we must apply at the outer
edge $r=r_{\rm max}$ of the spatial domain. It is, in fact, possible
to evolve the equations (\ref{alpha1}--\ref{v12}) numerically without
fixing boundary conditions for $\alpha_i$ and $\phi_i$ at the outer
end point of the spatial grid, provided we take $r=r_{\rm max}$ to be
sufficiently large that the total fluid density is always zero
there. However, any increase in the size of the spatial domain comes
at a heavy computational cost for a fixed spatial resolution, so we
choose to apply suitable boundary conditions at the outer end point of
the spatial grid.  To do so, we use the 4 rightmost end points of the
grid to calculate third-order spatial derivatives of $\alpha_i$ and
$\phi_i$ at $r=r_{\rm max}$ and thereby predict to third-order in $r$
the values of these variables at an `extra' endpoint, one grid step
beyond the desired boundary. Since such a scheme is accurate to third
order, it is sufficient for the consistent solution of our system of
second-order PDEs.

Having specified the above initial and boundary conditions on the
solution, the equations (\ref{alpha1}--\ref{v12}) can then be
integrated numerically. The most straightforward numerical approach is
to use a simple first-order time-stepping technique in which the
initial distributions $\alpha_{i}(r,0)$ and $\phi_{i}(r,0)$ are
`marched forward': at each time-step $t\mapsto t+\Delta t$, we
calculate the required spatial partial derivatives on the right-hand
sides of (\ref{alpha1}--\ref{phi2}) by taking differences and then add
$\Delta t\times\partial\alpha_{i}/\partial t$ and $\Delta
t\times\partial\phi_{i}/\partial t$ to $\alpha_{i}$ and $\phi_{i}$
(respectively) at each grid point. It is well-known, however, that
such a scheme is prone to numerical instability (a fact that we
unfortunately verified when we implemented it). We therefore instead
employed a fourth-order Runge--Kutta integration technique (see
e.g. Press et al. 2007), which calculates spatial derivatives (again
by simple differencing) at several points throughout the time interval
through which the solution is being advanced. This offers a
significant increase in accuracy over a standard Euler, or simple
finite differencing, technique, which calculates spatial derivatives
only at the beginning of the time interval.

\subsection{Toy physical system}

To demonstrate that our approach can indeed be used to describe the
evolution of a (spherically-symmetric) two-fluid system after the
fluids overlap, we apply it to a simple physical system similar to the
spherical overdensity model considered analytically in
Section~\ref{sec:sphericalmodel}. In that case, however, the inner and
outer fluid regions did not overlap, the outer fluid region was of
infinite spatial extent, and the density profiles possessed sharp
step-function discontinuties. We consider here instead a
modified physical system, as follows. First, we consider both inner
and outer fluid regions to be of finite spatial extent. Second, we
specify initial density and velocity distributions for both fluids
that are consistent with flat-universe evolution, since it is
straightforward to construct such a system in which overlap of fluids
is guaranteed to occur. Finally, since our numerical solution scheme
relies on obtaining accurate spatial derivatives, it is clearly
feasible only for initial density profiles that are smoothed to some
degree. Nonetheless, the smoothing is sufficiently slight that we can
compare the results we obtain numerically to known analytic results
for the evolution of an annulus of a flat universe, as found in
Section~\ref{sec:sphericalmodel}.

\begin{figure*}
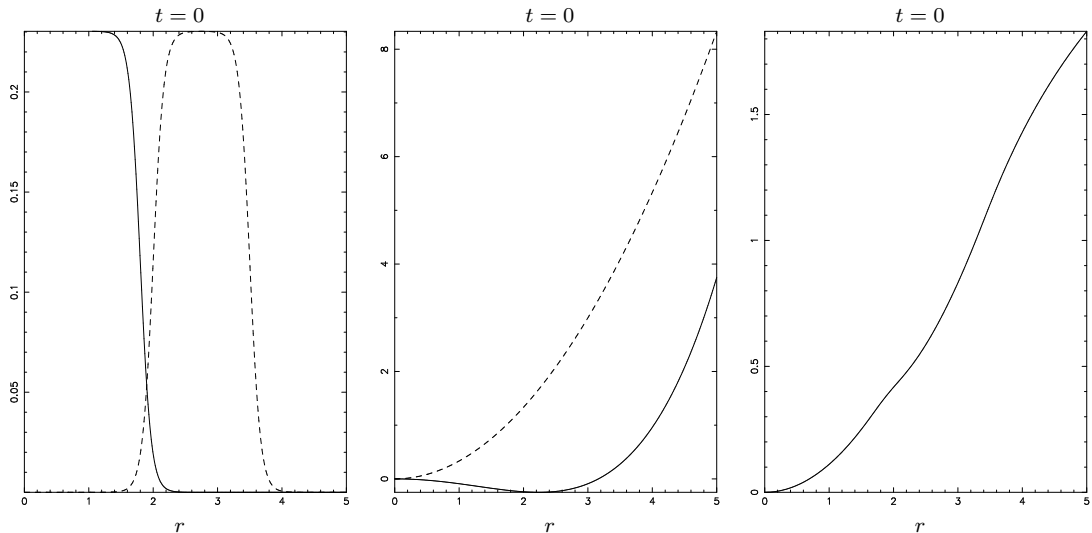

\begin{center}
\begin{tabular}{ccc}
$t=0$ & $t=0$ & $t=0$ \\
\includegraphics[width=45mm]{FINAL_PLOTS/output_plotinitialalpha.ps} & 
\includegraphics[width=45mm]{FINAL_PLOTS/output_plotinitialphi.ps} & 
\includegraphics[width=45mm]{FINAL_PLOTS/output_plotinitialV.ps} \\
$r$ & $r$ & $r$  
\end{tabular}
\caption{The initial square-root densities, $\alpha_1$ and $\alpha_2$
  (left), initial velocity potentials, $\phi_1$ and $\phi_2$ (middle)
  and the initial gravitational potential $V$ (right) for the
  numerical integration of the spherically-symmetric two-fluid system.
  The solid lines refer to the fluid 1 and the dashed lines to fluid 2.
  The outer fluid (fluid 2) is given a small inwards initial velocity,
  as shown.}
\label{initialplots}
\end{center}
\end{figure*}

The assumed initial $\alpha_i$ and $\phi_i$ distributions for each
fluid are shown in Fig.~\ref{initialplots}, together with the
resulting initial gravitational potential $V$ derived from
(\ref{vdint}) using a single application of the Simpson's rule method
described above.
The initial $\alpha_i$ profiles are step functions smoothed with a
relatively narrow $\tanh$ function. The initial profile $\phi_1$ for
the inner fluid is that of a flat universe, i.e. quadratic in $r$, and
the velocity potential for the outer fluid was taken to be
$\phi_2(r,0)=0.01 - 0.1r^2$.  We set the outer boundary at $r_{\rm
  max}=5$, and used $401$ grid points to sample the spatial
domain. The equations were evolved over 651 time steps, with a total
integration time of 3.5. It was checked that the ratio of spatial and
time resolutions $\Delta x/\Delta t$ remained greater than or equal to
the greatest fluid velocity in the system throughout the entire
integration time. 

It is worth mentioning how the accuracy of the numerical integration
is affected by the choice of the initial conditions. As expected, very
sharp edges on the $\alpha_i$ profiles lead to inaccuracies in the
calculations spatial derivatives via finite differencing when the
`width' of the profile edges is close to the spatial grid-spacing;
these inaccuracies can then propagate into the rest of the spatial
domain. Also, if either fluid velocity is too fast at any point, such
that it exceeds $\Delta x/\Delta t$, then inaccuracies again
propagate through the numerical solutions. This may occur directly by
setting the initial velocity too high, but high velocities can also
occur later in the evolution if the initial density of either fluid is
set very large, resulting in faster gravitational collapse.

\begin{figure*}
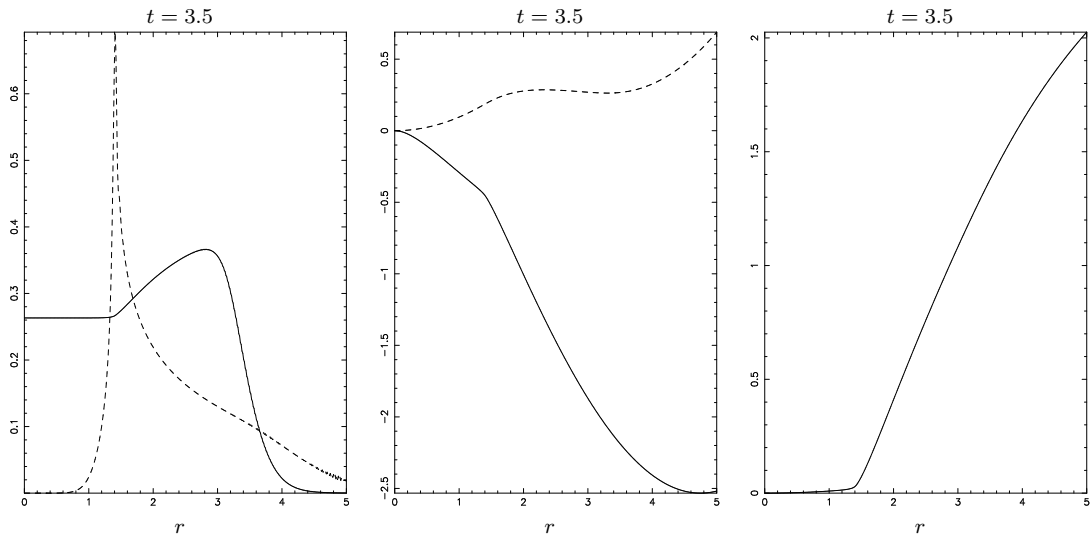

\begin{center}
\begin{tabular}{ccc}
$t=3.5$ & $t=3.5$ & $t=3.5$ \\
\includegraphics[width=45mm]{FINAL_PLOTS/output_plotalpha.ps} & 
\includegraphics[width=45mm]{FINAL_PLOTS/output_plotphi.ps} & 
\includegraphics[width=45mm]{FINAL_PLOTS/output_plotV.ps} \\
$r$ & $r$ & $r$  
\end{tabular}
\caption{As for Fig.~\ref{initialplots}, but 
after evolving the system to a time $t=3.5$.}
\label{finalplots}
\end{center}
\end{figure*}

Given the initial conditions illustrated in Fig.~\ref{initialplots},
the resulting $\alpha_i$ and $\phi_i$ distributions for each fluid
after evolving the system to $t=3.5$, together with resulting
gravitational potential $V$, are shown in Fig.~\ref{finalplots}.
We see that, as the two fluids overlap, new features develop in the
profiles. In particular, we note a pronounced caustic in fluid 2 at a
finite distance from the origin, which signals that fluid is `piling
up' at this radius as a result of shell-crossing. Fluid 1 also possess
a corresponding `hump' at radii larger than the position of the
caustic in fluid 2.

Finally, to check our code, we also considered a model in which the
initial density distribution of fluid 1 was set to zero everywhere,
but the initial density of fluid 2 has the same profile as in
Fig.~\ref{initialplots} (left panel) and its velocity potential was
taken to be that of a spatially-flat universe.  In this case, aside
from minor differences resulting from the smooth edges of the density
distribution of fluid 2, we expect this fluid to evolve as (an annulus
of) a spatially-flat universe (with $\Lambda=0$), for which the
analytic solution is given by (\ref{flatsolution1}) and
(\ref{flatsolution2}). Our numerical results were found to agree well
with this analytic solution. This example also demonstrates that,
because of the (slight) smoothing of the initial density profile of
fluid 2, the total density at the origin $r=0$ was non-zero to machine
precision, thereby satisfying our earlier assumption in solving
equation (\ref{vdint}) for the gravitational potential. Hence our
scheme can still be used in many cases to evolve systems which
`appear' to have a hole at the centre.

\section{Conclusions}\label{sec:conc}

We have considered the use of the \schro formalism for fluid dynamics
to model the evolution of cosmological dark matter. Our approach
differs from previous work in this area in that we consider the full
non-linear Schr\"odinger--Poisson (SP) system, without making any
approximations. In particular, we retain the `quantum pressure' term,
which previous authors have discarded to regain a linear
wave-equation. By so doing previous authors have had to work in the
so-called correspondence limit $\nu \to 0$, where $\nu$ is a parameter
that changes the spatial and velocity resolution of the solutions.

We have shown that cosmological solutions to the full SP system can be
readily obtained for closed and spatially-flat universes.  Moreover,
these solutions can be combined to obtain a global solution to the SP
system for a spherical overdensity. This necessitated determining the
appropriate boundary conditions on the solutions for each homogeneous
region of the model, which differ from those usually assumed in
quantum mechanics. We also highlight that the issue of what
conditions to apply to wavefunctions at a moving boundary remains
a matter of debate in the quantum mechanics literature.

We consider how the \schro formalism for classical fluids can be
extended to multiple fluids and derive an appropriate
wave-equation for the two-fluid case, which could be easily
generalised to more fluids. In particular, we find that for
consistency of the equations the multi-fluid wavefunction must be
factorisable into wavefunctions for each fluid separately; `entangled'
states appear to be problematic. Nonetheless, our approach provides a
new and consistent method for dealing with multiple fluids.  We
illustrate our method for describing multi-streaming fluids by
applying it to the evolution of a modified spherical overdensity toy
model, in which the two fluids eventually overlap. We integrate the SP
equations numerically to obtain physically sensible results, which
illustrate the formation of cusps.

We consider our investigations to show that the full non-linear
SP system has the potential to be a useful tool for theoretical
investigations of cosmological fluid evolution.  The non-linearity of
the equations can be troublesome for analytical investigations
(although not always), but poses few problems for following the fluid
evolution numerically.  In any case, we believe these difficulties are
offset by the advantages it offers in
ensuring that the fluid density is always positive. In future work, we
plan to investigate linear perturbation methods within the \schro
formalism to determine whether such an approach offers any advantages
over traditional first-order perturbation methods applied to the standard
fluid equations in following cosmological structure formation in the
linear regime.

\subsection*{ACKNOWLEDGMENTS}
RJ acknowledges financial support from the Gates Cambridge Trust.

\bsp 
\label{lastpage}
\end{document}